# Make Scientific Conferences More Inclusive by Increasing Transparency in Speaker Selection

Tibor Szilvási

tszilvasi@ua.edu

Department of Chemical and Biological Engineering, The University of Alabama, Tuscaloosa, AL 35487, United States

Scientific conferences play an important part in every scientist's life. We can finally meet our peers, call attention to our research and ourselves, and disseminate interesting results. Being an invited/plenary speaker is considered a clear sign of excellence and thus several universities in multiple countries weigh the number of invited/plenary lectures when a candidate is evaluated for tenure. In addition, invited/plenary speakers often get financial aid to attend the conference that can often be a deal breaker for scientists working in countries where science is severely underfinanced. Therefore, there is a clear need to distribute conference invitations that are as inclusive and unbiased as possible.

Here, I suggest an approach that can mitigate biases and help make scientific conferences more inclusive for everybody by increasing the transparency in speaker selection. The core idea is that the scientific organizing committee should define the criteria for excellence that is openly available to the community and then a certain fraction of invited speakers is selected randomly from the pool of candidates who met the criteria. I emphasize that scientific organizing committees have always had to define excellence implicitly during their selection process; thus, my suggestion is only to make it well-defined and public. Such criteria of excellence can be, for example, X number of corresponding author papers in a set of journals relevant to the topical area in the last Y years, or a combination of multiple factors that forms a scoring system such that candidates enter the random selection process with a different probability to be selected. It does not matter what specific selection criterion is chosen, the main point is that it is to be transparent and predetermined. After the criteria become public, the scientific organizing committee and the community can propose candidates via nomination or self-nomination for random selection. When the random selection process is done the committee can release the names of invited speakers and those who have met the criteria but were not selected, which can provide feedback to the community and candidates that the selection process was broad and inclusive (e.g., criteria were not tailored to a few candidates). Randomized selection does have drawbacks; for instance, it cannot efficiently capture newly emerging subdisciplines. Thus, it is logical that the scientific organizing committee decides on some invited speakers to handle potential drawbacks. Nevertheless, these invited speakers should still fulfill the predetermined selection criteria if possible and their special selection circumstances should be also disclosed.

I do not think that the proposed selection process is perfect (in fact, there is no such method) but I do think that it is more transparent than the current practice and transparency will eventually lead to more inclusivity and less bias. In addition, I can list several important advantages (without particular order of importance):

1) Forces communities to discuss openly what excellence means for them. I do not expect that excellence can be defined in a way that everybody will agree on, but I think a set of minimum criteria can be obtained that is sufficient for the partial randomized selection process. Open definition of excellence helps newcomers set the bar, and it makes sure they are treated fairly even if they are not part of the inner circles.

2) Can be continuously improved by the entire community. Open information on the criteria of excellence and the list of considered speakers provides a means of detailed feedback from the entire community that other conference organizers can utilize improving their selection process.

3) Is an inclusive approach for everyone. Gender and geographic disparities are often brought up but there are other less obvious inclusivity issues. The randomized selection provides a proportional representation to every group even for those whose voice has not been heard.

4) Sends a positive message to both the selected and the non-selected researchers. E.g., female scientists often express their feeling that they are invited to fulfill an unofficial female quota so that the organizers can avoid criticism. Publicly disclosed criteria of excellence ensure that every speaker is selected based on merit. Disclosing the list of candidates based on the criteria of excellence also sends a positive message to non-selected researchers that their work is excellent, they were considered seriously, only the limited number of speaker slots resulted in the negative decision.

5) Frees the scientific organizing committees from accusations of partiality. The process and the information presented to the community is evidence that the committee has been working according to the highest standards.

As every change, the proposed process can create aversions. I envision that several researchers may find randomness unscientific, but I see it as a measurement that has an error bar and within error bar decisions can only be made based on randomness or human bias. I also mention that I am not the first one to suggest or even implement such methods to eliminate human bias. For example, the Swiss National Science Foundation has started using lottery drawing to eliminate human bias when choosing between applications of similar quality.[1] Since 2013, the Health Research Council of New Zealand has allocated roughly 2% of its annual funding using lotteries to decide between projects reviewers suggested for funding.[1]

I also foresee that scientific organizing committees can be against the suggestion because transparency is an additional burden and defining criteria of excellence for the first time can be challenging and may trigger harsh criticism if it is not done well. Pressure from leaders of scientific organizations overseeing conferences, regularly invited speakers, and the wider community will likely be needed to force the initiative. Discussion in the wider community on excellence and patience towards organizers implementing it well will be also required to make the proposed methods successful.

References:

1. https://www.nature.com/articles/d41586-021-01232-3